\documentclass[10pt,twocolumn,letterpaper]{article}

\usepackage{acpr}
\usepackage{times}
\usepackage{epsfig}
\usepackage{graphicx}
\usepackage{amsmath}
\usepackage{amssymb}
\usepackage{array}

\usepackage[pagebackref=true,breaklinks=true,letterpaper=true,colorlinks,bookmarks=false]{hyperref}

\acprfinalcopy 

\def\acprPaperID{308} 

\ifacprfinal\fi
\begin{document}

\title{Fine-Grain Annotation of Cricket Videos}

\author{Rahul Anand Sharma\\
CVIT, IIIT-Hyderabad\\
Hyderabad, India\\
{\tt\small rahul.anand@research.iiit.ac.in}
\and
Pramod Sankar K.\\
Xerox Research Center India\\
Bengaluru, India\\
{\tt\small pramod.kompalli@xerox.com}
\and
C. V. Jawahar\\
CVIT, IIIT-Hyderabad\\
Hyderabad, India\\
{\tt\small jawahar@iiit.ac.in}
}

\maketitle

\begin{abstract}
The recognition of human activities is one of the key problems in video understanding. 
Action recognition is challenging even for specific categories of videos, such as sports, that contain only a small
set of actions. 
Interestingly, sports videos are accompanied by detailed commentaries available online, which could be used to perform action annotation in a weakly-supervised setting. 
For the specific case of Cricket videos, we address the challenge of temporal segmentation and annotation of actions with semantic descriptions.
Our solution consists of two stages. In the first stage, the video is segmented into ``scenes'', by utilizing the scene category information extracted from text-commentary. 
The second stage consists of classifying video-shots as well as the phrases in the textual  description into various
categories. The relevant phrases are then suitably mapped to the video-shots. The novel aspect of this work is the fine temporal scale at which semantic information is assigned to the video. As a result of our approach, we enable 
retrieval of specific actions that last only a few seconds, from several  hours of video. 
This solution yields a large number of labelled exemplars, with no manual effort, that could be used by 
machine learning algorithms to learn complex actions.

\end{abstract}

\section{Introduction}

The labeling of human actions in videos is a challenging problem for computer vision systems.
There are three difficult tasks that need to  be solved to perform action recognition: 1) identification of 
the sequence of frames that involve an action performed, 2) localisation of the person performing the action 
and 3) recognition of the pixel information to assign a semantic label. While each of these tasks could be
solved independently, there are few robust solutions for their joint inference in generic videos. 

Certain categories of videos, such as Movies, News feeds, Sports videos, etc. contain domain specific cues that could be exploited  towards better understanding of the visual content. 
For example, the appearance of a Basketball court~\cite{swearscvpr14}
 could help in locating and tracking players and their movements. 
However, current visual recognition solutions have only seen limited success towards fine-grain 
action classification. For example, it is difficult to automatically distinguish a ``forehand'' from a 
``half-volley'' in Tennis. Further, automatic generation of semantic descriptions is a much harder 
task,  with only limited success in the image domain~\cite{karpathycvpr15}.

\begin{figure}
\begin{tabular}{m{8cm}}

\includegraphics[width=2.6cm]{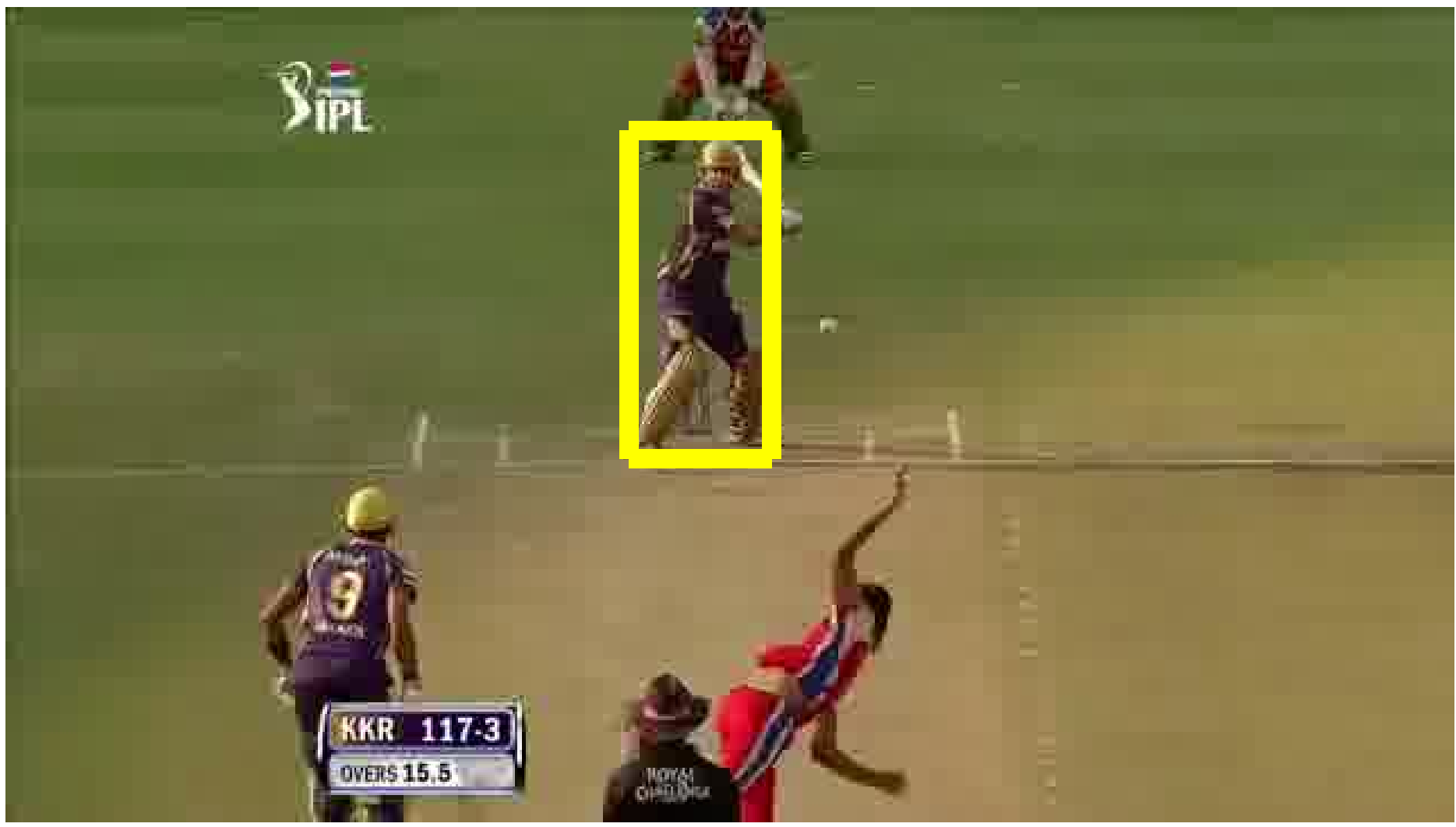}
\includegraphics[width=2.6cm]{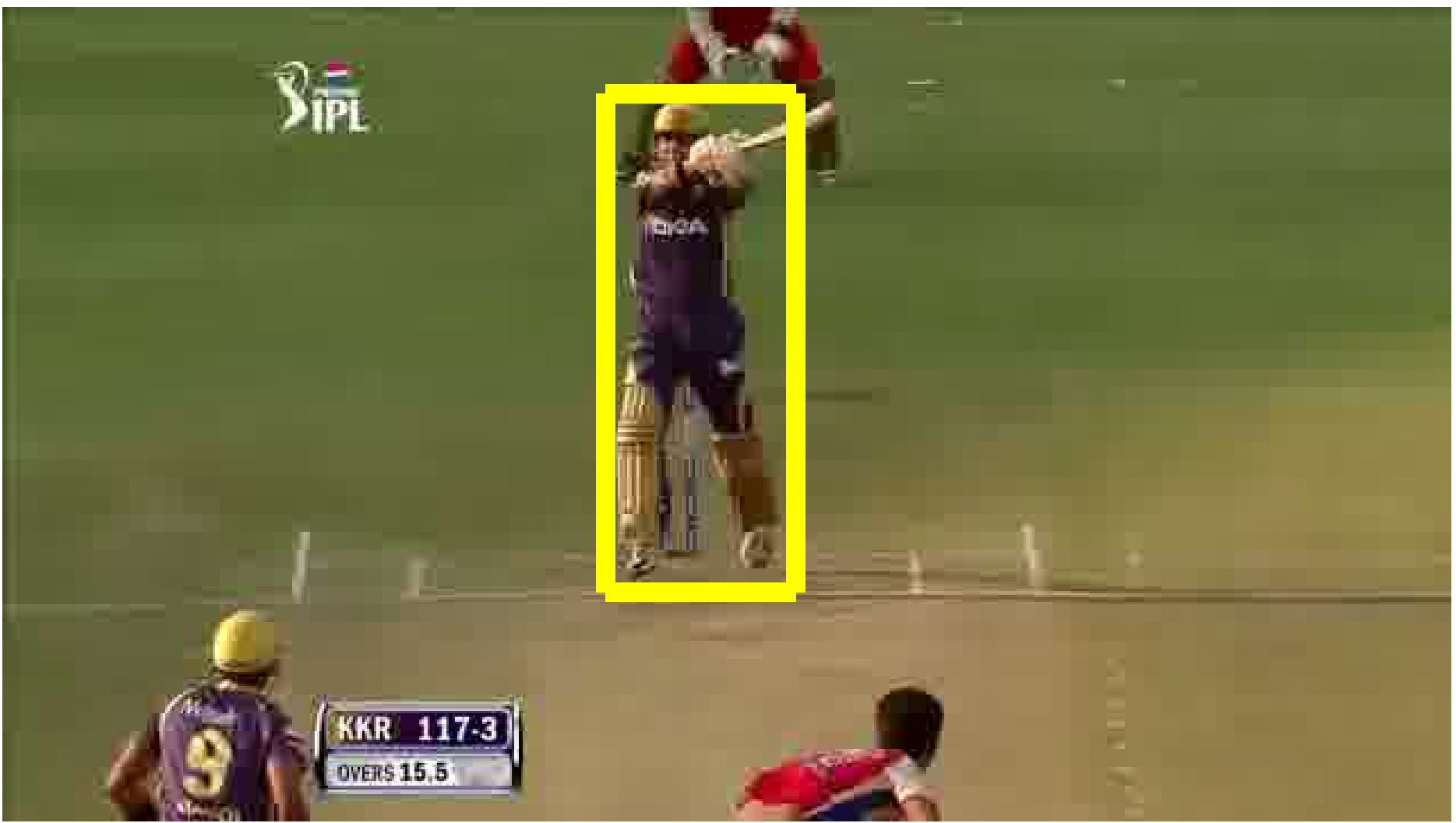}
\includegraphics[width=2.6cm]{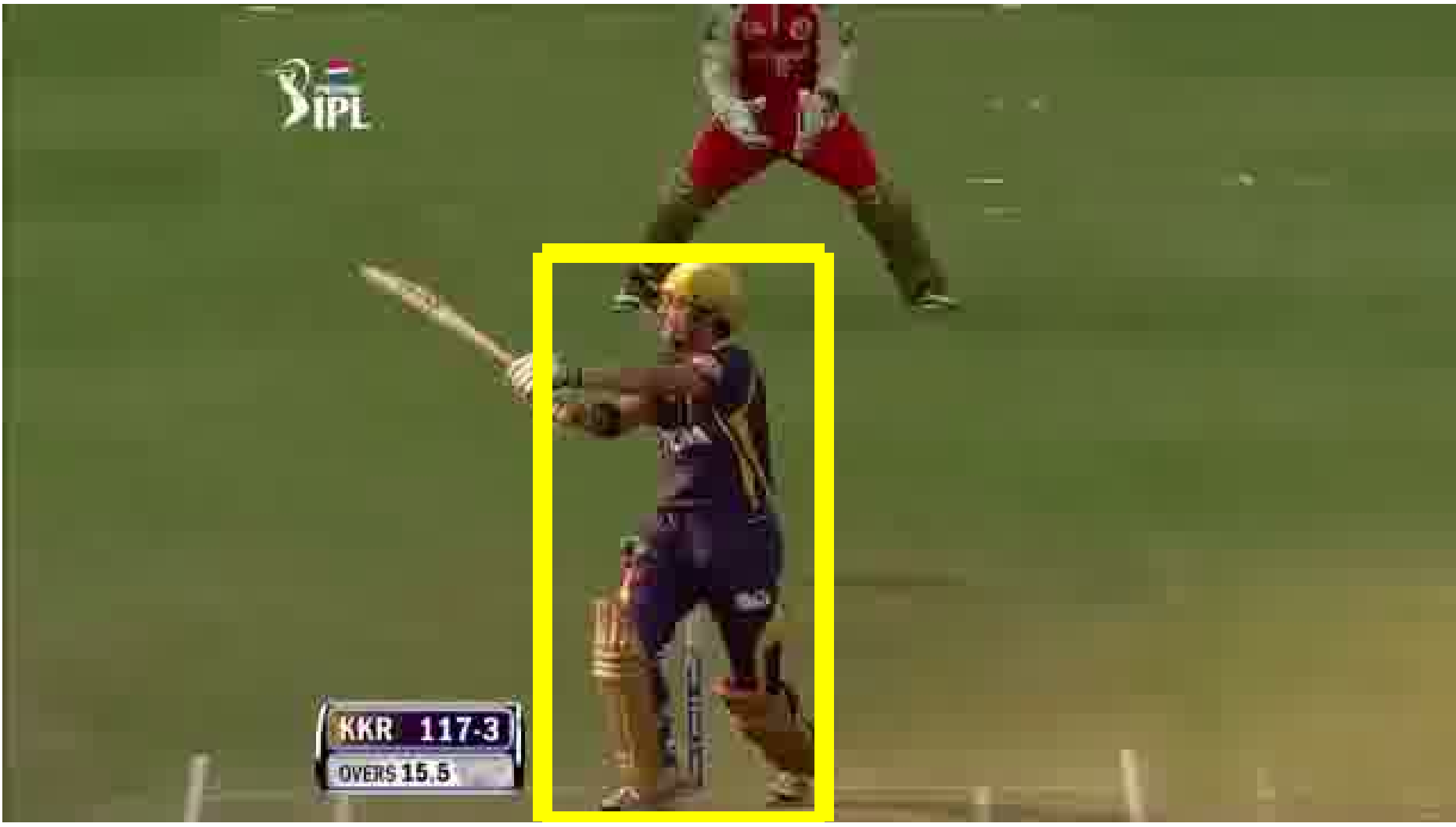}\\
\hline
Batsman: \em Gambhir \\
\hline
Description: \emph{``he pulls it from outside off stump and just manages to clear the deep square leg rope''} \\
\hline
\end{tabular}
\caption{The goal of this work is to annotate Cricket videos with semantic descriptions at a fine-grain spatio-temporal scale. In this example, the batsman action of a ``pull-shot'', a particular manner of hitting the ball, is labelled accurately as a result of our solution. Such a semantic description is impossible to obtain using current visual-recognition techniques alone. The action shown here lasts a mere 35 frames (1.2 seconds).}
\label{fig:motiv}
\vspace{-0.4cm}
\end{figure}

Instead of addressing the problem using visual analysis alone, several researchers proposed to 
utilize relevant \emph{parallel} information to build better solutions~\cite{Everingham06a}. 
For example, the scripts available for movies provide a weak supervision to perform 
person~\cite{Everingham06a} and action recognition~\cite{laptev08}.
Similar parallel text in sports was previously used to detect events in soccer videos, and index 
them for efficient retrieval \cite{webcast2,zhangmir07}. Gupta~\emph{et al.}~\cite{guptacvpr09} 
learn a graphical model from annotated baseball videos that could then be used to generate captions 
automatically. Their generated captions, however, are not semantically rich. 
Lu~\emph{et al.}~\cite{lupami13} show that the weak supervision from parallel text results in superior 
player identification and tracking in Basketball videos.

\begin{figure*}[t]
\center{\includegraphics[width=17cm]{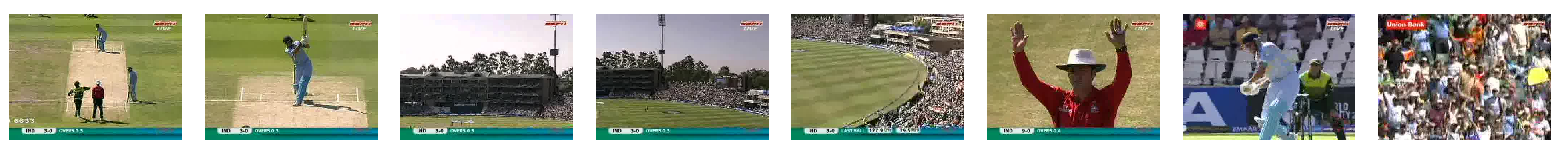}}
\begin{tabular} {p{1.8cm} p{2cm} p{5.75cm} p{1.55cm} p{2.15cm} p{2cm}}
\centering Bowler & \centering Batsman & \centering Ball-in-the-air & \centering Umpire  & \centering Replay &  Crowd  \\
\centering Run-Up  & \centering Stroke & \centering (Sky)  & \centering Signal  & & Reaction \\
\end{tabular}
\caption{Typical visuals observed in a \emph{scene} of a Cricket match. Each event begins with the \emph{Bowler} running to throw the ball, which is hit by the \emph{Batsman}. The event unfolds depending on the batsman's stroke. The outcome of this particular scene is \emph{6-Runs}. While the first few shots contain the real action, the rest of the visuals have little value in post-hoc browsing.} 
\label{fig:crickevents}
\vspace{-0.25cm}
\end{figure*}

In this work, we aim to label the actions of players in \emph{Cricket} videos using parallel information in the form of online text-commentaries~\cite{cricinfo}. The goal is to label the video at the shot-level with the semantic descriptions of actions and activities. Two challenges need to be addressed towards this goal. Firstly, the visual
and textual modalities are not aligned, i.e. we are given a few pages of text for a four hour video with no other
synchronisation information. Secondly, the text-commentaries are very descriptive, where they assume that the
person reading the commentary understands the keywords being used. Bridging this semantic gap over video data
is much tougher than, for example, images and object categories. 


\begin{figure}[t]

\center{\framebox{\includegraphics[width = 8cm]{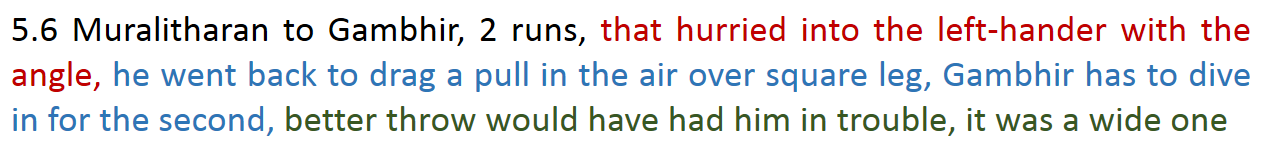}}}
\caption{An example snippet of commentary obtained from Cricinfo.com. The commentary follows the format: event number and player involved along with the outcome (2 runs). Following this is the descriptive commentary: (red) bowler actions, (blue) batsman action and (green) other player actions.} 
\label{fig:commentary}
\vspace{-0.4cm}
\end{figure}

\subsection{Our Solution}

We present a two-stage solution for the problem of fine-grained Cricket video segmentation and annotation. 
In the first stage, the goal is to \emph{align} the two modalities at a ``scene'' level.
This stage consists of a joint synchronisation and segmentation of the video with the text commentary. 
Each scene is a meaningful event that is a few minutes long (Figure~\ref{fig:crickevents}), and described by a small set of sentences in the commentary (Figure~\ref{fig:commentary}). 
The solution for this stage is  inspired from the approach proposed in~\cite{cricksegicvgip06}, and presented in Section~\ref{sec:seg}.

Given the scene segmentation and the description for each scene, the next step is to align the individual descriptions
with their corresponding visuals.
At this stage, the alignment is performed between the video-shots and \emph{phrases} of the text commentary.
This is achieved by classifying video-shots and phrases into a known set of categories, which allows them to 
be mapped easily across the modalities, as described in Section~\ref{sec:shotphra}.
As an outcome of this step, we could obtain fine-grain annotation of player actions, such as those presented in Figure~\ref{fig:motiv}.

Our experiments, detailed in Section~\ref{sec:expts} demonstrate that the proposed solution is sufficiently 
reliable to address this seemingly challenging task. The annotation of the videos allow us to build a retrieval system that can search across hundreds of hours of content for specific actions that last only a few seconds.
Moreover, as a consquence of this work, we generate a large set of fine-grained labelled videos, that could be
used to train action recognition solutions.

\begin{figure*}[t]

\center{\framebox{\includegraphics[width=6cm,height=4cm]{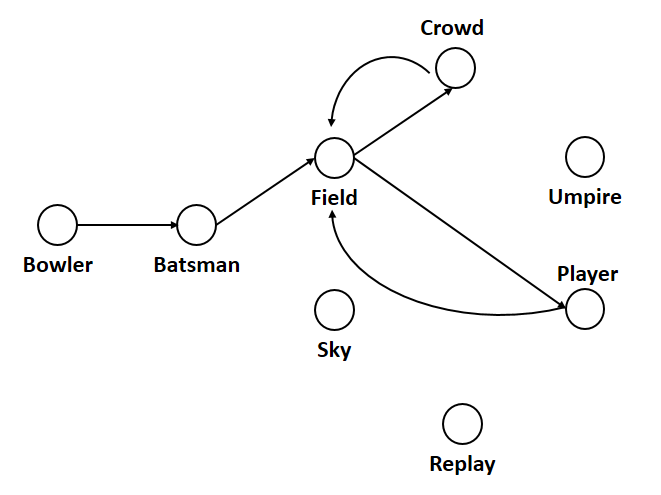}}
\framebox{\includegraphics[width=6cm,height=4cm]{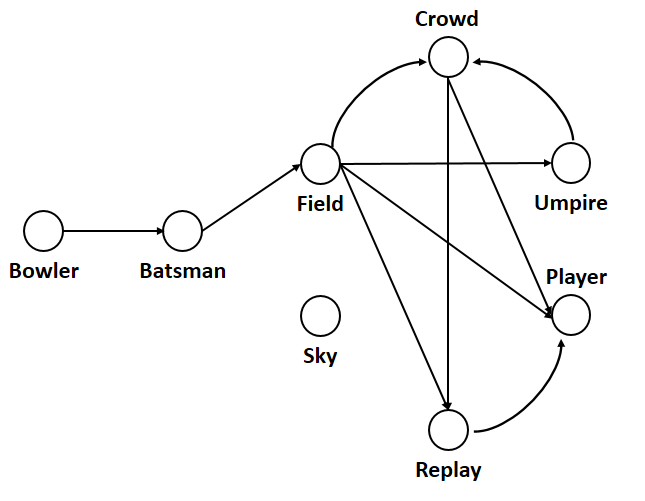}}}
\caption{State transition diagrams for two scene categories: (left) One Run and (right) Four-Runs. Each shot is classified into one of the given states. Only the prominent state-transitions are shown, each transition is associated with a probability (not shown for brevity). Notice how the one-run scene includes only a few states and transitions, while the four-run model involves a variety of visuals. However, the ``sky'' state is rarely visited in a Four, but is typically seen in Six-Runs and Out models.} 
\label{fig:scenemodels}
\vspace{-0.5cm}
\end{figure*}

\section{Scene Segmentation}
\label{sec:seg}

A typical scene in a Cricket match follows the sequence of events depicted in Figure~\ref{fig:crickevents}.
A scene always begins with the \emph{bowler} (similar to a pitcher in Baseball) running towards and throwing the ball at the \emph{batsman}, who then plays his \emph{stroke}.
The events that follow vary depending on this action.
Each such scene is described in the text-commentary as shown in Figure~\ref{fig:commentary}. 
The commentary consists of the event number, which is not a time-stamp; the player names, which are hard to
recognise; and detailed descriptions that are hard to automatically interpret. 

It was observed in~\cite{cricksegicvgip06} that the visual-temporal patterns of the scenes are conditioned on the \emph{outcome} of the event.
In other words, a \emph{1-Run} outcome is visually different from a \emph{4-Run} outcome. 
This can be observed from the state-transition diagrams in Figure~\ref{fig:scenemodels}.
In these diagrams,  the \emph{shots} of the video are represented by visual categories such as 
\emph{ground}, \emph{sky}, \emph{play-area}, \emph{players}, etc. 
The sequence of the shot-categories is represented  by the arrows across these states. 
One can observe that for a typical Four-Runs video, the number of shots and their transitions are lot more complex than that of a 1-Run video.
Several shot classes such as \emph{replay} are typically absent for a 1-Run scene, while a replay is expected as the third or fourth shot in a Four-Runs scene.

While the visual model described above could be useful in recognizing the scene category for a given video segment, it cannot be immediately used to \emph{spot} the scene in a full-length video. 
Conversely, the temporal segmentation of the video is not possible without using an appropriate model for the individual scenes themselves. This chicken-and-egg problem can be solved by utilizing the scene-category information from the parallel text-commentary. 

Let us say $F_i$ represents one of the $N$ frames and $S_k$ represents the category of  the $k$-th scene.
The goal of the scene segmentation is to identify anchor frames $F_i, F_j$, which 
are most likely to contain the scene $S_k$.
The optimal segmentation of the video can be defined by the recursive function
\[
C(F_i, S_k) = \max_{j \in [i+1, N]} \{ p(S_k \ | \  [F_i, F_j]) + C(F_{j+1}, S_{k+1}) \},
\]
where $ p(S_k \ | \  [F_i, F_j])$ is the probability that the sequence of frames $[F_i,F_j]$ belong to the scene category $S_k$.
This probability is computed by matching the learnt scene models with the sequence of 
features for the given frame set. 
The optimisation function could be solved using Dynamic Programming (DP). 
The optimal solution is found by backtracking the DP matrix, which provides 
the scene anchor points $F_{S_1}, F_{S_2}, ..., F_{S_K}$.


We thus obtain a temporal segmentation of the given video into its individual  scenes. 
A typical segmentation covering five \emph{overs}, is shown in Figure~\ref{fig:sceneseg}.
The descriptive commentary from the parallel-text could be used to annotate the scenes, for text-based search
and retrieval.
In this work, we would like to further annotate the videos at a much finer temporal scale than the scenes, i.e., we would like to
annotate at the shot-level. 
The solution towards shot-level annotation is presented in the following Section.

\section{Shot/Phrase Alignment}
\label{sec:shotphra}

Following the scene segmentation, we obtain an alignment between minute-long clips with a paragraph of text.
To perform fine-grained annotation of the video, we must segment both the video clip and the descriptive text.
Firstly, the video segments at scene level are over-segmented into shots, to ensure that it is unlikely to map multiple actions into the same shot.
In the case of the text, given that the commentary is free-flowing, the action descriptions need to be identified at a finer-grain than the sentence level. Hence we choose to operate at the phrase-level, by segmenting sentences at all punctuation marks. 
Both the video-shots and the phrases are classified into one of these categories: \emph{\{Bowler Action, Batsman Action, Others\}}, by learning suitable classifiers for each modality. Following this, the individual phrases could be mapped to the video-shots that belong to the same category.

\subsection{Video-Shot Recognition}

In order to ensure that the video-shots are \emph{atomic} to an action/activity, we perform an over-segmentation of the video. We use a window-based shot detection scheme that works as follows. For each frame $F_i$, we compute its difference with every other frame $F_j$, where $j \in [i-w/2, i+w/2]$, for a chosen window size $w$, centered on $F_i$.  If the maximum frame difference within this window is greater than a particular threshold $\tau$, we declare $F_i$ as a shot-boundary. We choose a small value for $\tau$  to ensure over-segmentation of scenes. 

Each shot is now represented with the classical Bag of Visual Words (BoW) approach~\cite{sivic03}. SIFT features are first computed for each frame independently, which are then clustered using the K-means clustering algorithm to build a visual vocabulary (where each cluster center corresponds to a visual word). Each frame is then represented by the normalized count of number of SIFT features assigned to each cluster (BoW histograms). The shot is represented by the average BoW histogram over all frames present in the shot. The shots are then classified into one of these classes: \emph{\{Bowler Runup,  Batsman Stroke, Player Close-Up, Umpire, Ground, Crowd, Animations, Miscellaneous\}}. The classification is performed using a multiclass Kernel-SVM. 

The individual shot-classification results could be further refined by taking into account the temporal neighbourhood information. Given the strong structure of a Cricket match, the visuals do not change arbitrarily, but are predictable according to the sequence of events in the scenes. Such a sequence could be modelled as a Linear Chain Conditional Random Field (LC-CRF)~\cite{laffertyicml01}. 
The LC-CRF, consists of nodes corresponding to each shot, with edges connecting each node with its previous and next node, resulting in a linear chain.
The goal of the CRF is to model $P(y_1, ..., y_n | x_1, ..., x_n) $, where $x_i$ and $y_i$ are the input and output sequences respectively.
The LC-CRF is posed as the objective function 
\[
p(Y\|X) = \exp(\sum_{k=1}^{K} a_u(y_k) + \sum_{k=1}^{K-1}a_p(y_k,y_{k+1})     )/Z(X)
\]
Here, the unary term is given by the class-probabilities produced by the shot classifier, defined as 
\[a_{u}(y_k) = 1 - P(y_k | x_i).\]
The pair-wise term encodes the probability of transitioning from a class $y_{k}$ to $y_{k+1}$ as 
\[a_p(y_k,y_{k+1}) = 1 - P(y_{k+1} | y_{k}).\]
The function $Z(X)$ is a normalisation factor.
The transition probabilities between all pairs of classes are learnt from a training set of labelled videos. 
The inference of the CRF is performed using the forward-backward algorithm.

\begin{figure}
\centering 
\framebox{\includegraphics[width=7.5cm]{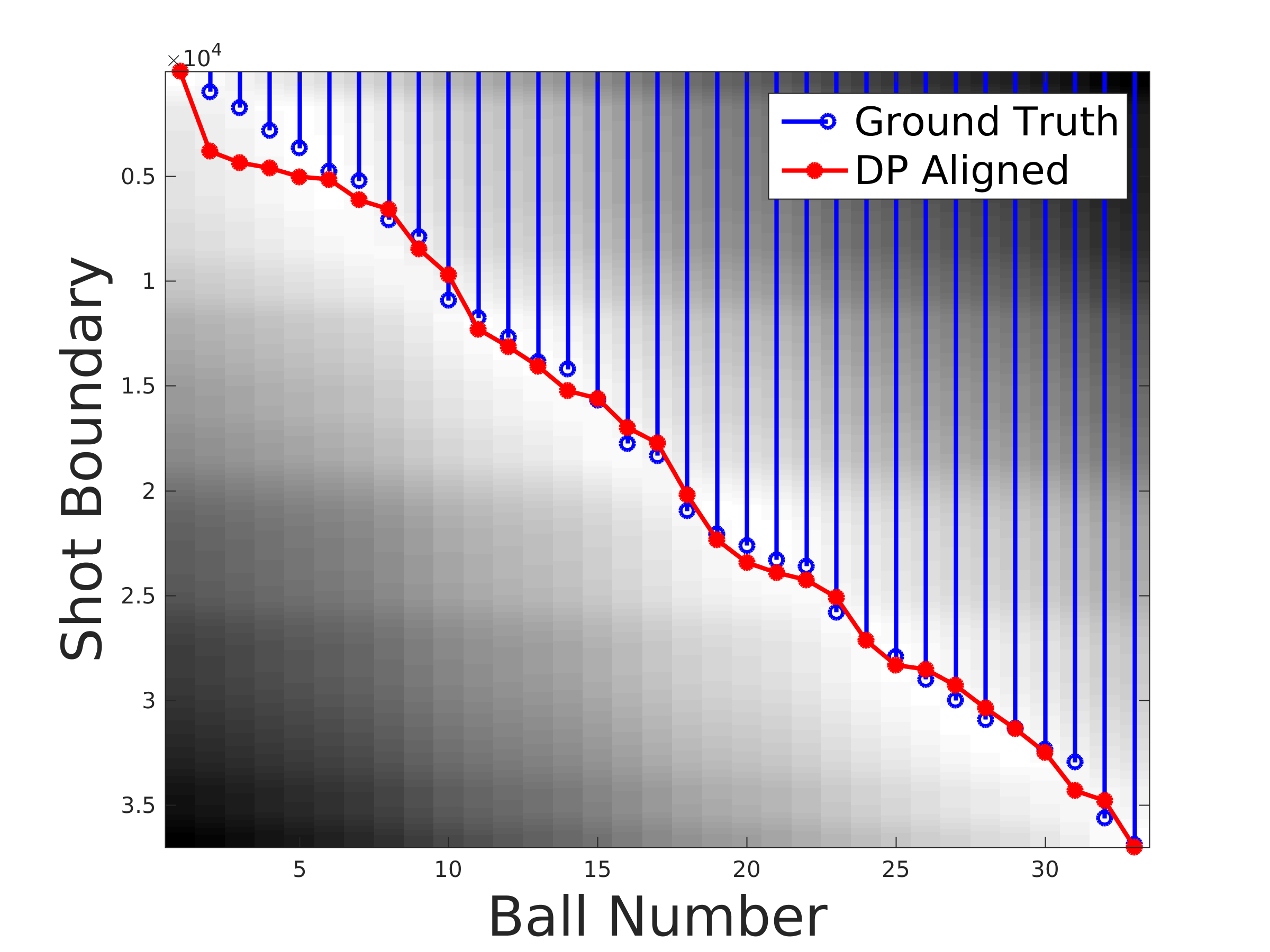}}

\caption{Results of Scene Segmentation depicted over five-\emph{overs}. The background of the image  is the cost function of the scene segmentation. The optimal backtrack path is given as the red line, with the inferred scene boundaries marked on this path. The groundtruth segmentation of the video is given as the blue lines. It can be noticed that for most scenes the inferred shot boundary is only a few shots away from the groundtruth.} 
\label{fig:sceneseg}
\vspace{-0.5cm}
\end{figure}

\subsection{Text Classification}

The phrase classifier is learnt entirely automatically. 
We begin by crawling the web for commentaries of about 300 matches and segmenting the text into phrases.
It was observed that the name of the bowler or the batsman is sometimes included in the description, for example, ``\emph{Sachin} hooks the ball to square-leg''.
These phrases can accordingly be labelled as belonging to the actions of the Bowler or the Batsman.
From the 300 match commentaries, we obtain about 1500 phrases for bowler actions and about 6000 phrases for the batsman shot. 
We remove the names of the respective players and represent each phrase as a histogram of its constituent word occurrences. 
A Linear-SVM is now learnt for the bowler and batsman categories over this bag-of-words representation. 
Given a test phrase, the SVM provides a confidence for it to belong to either of the two classes. 

The text classification module is evaluated using 2-fold cross validation over the 7500 phrase dataset.
We obtain a recognition accuracy of {\bf 89.09\%} for phrases assigned to the right class of bowler or batsman action.

\section{Experiments}
\label{sec:expts}

\paragraph{Dataset: }
Our dataset is collected from the YouTube channel for the Indian Premier League(IPL) tournament. The format for this tournament is 20-overs per side, which results in about 120 scenes or events for each team. 
The dataset consists of 16 matches, amounting to about 64 hours of footage.
Four matches were groundtruthed manually at the scene and shot level, two of which are used for training and the other two for testing.

\begin{figure}
\begin{tabular}{|m{8cm}|}
\hline
\vspace{0.1cm}
\includegraphics[width = 1.9cm]{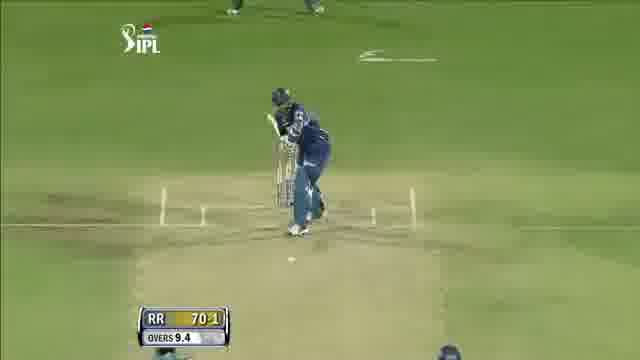} 
\includegraphics[width = 1.9cm]{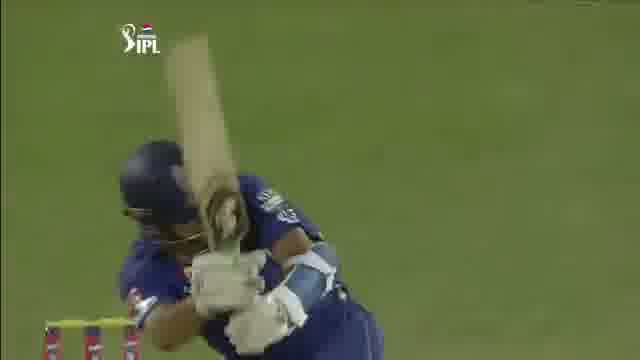} 
\includegraphics[width = 1.9cm]{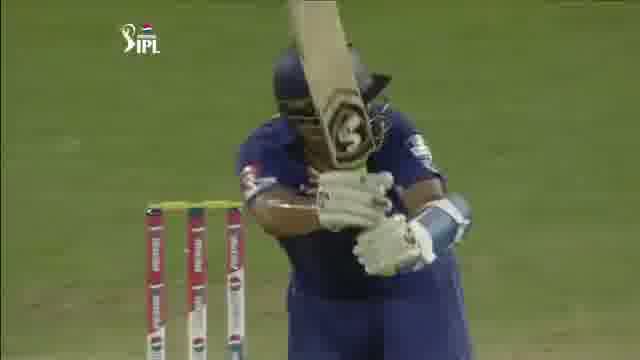} 
\includegraphics[width = 1.9cm]{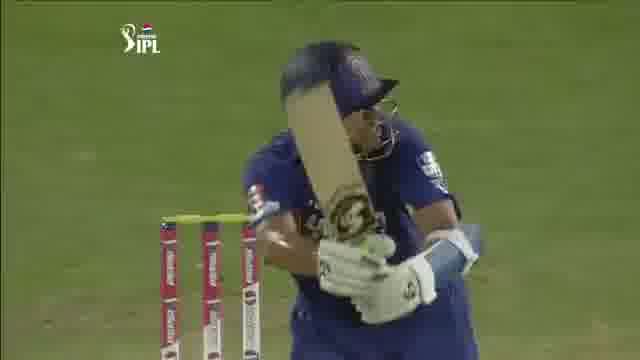} \\
\emph{on the pads once more Dravid looks to nudge it to square leg off the pads for a leg-bye} \\
\hline
\vspace{0.1cm}
\includegraphics[width = 1.9cm]{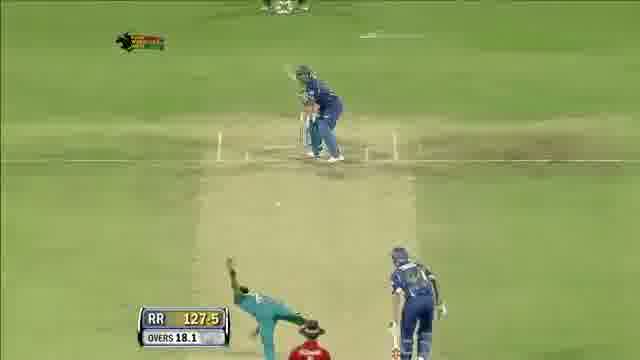} 
\includegraphics[width = 1.9cm]{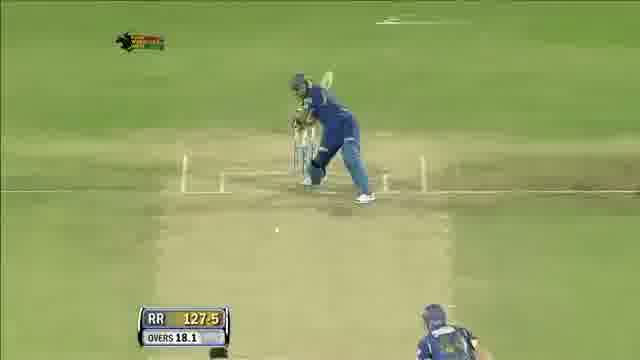} 
\includegraphics[width = 1.9cm]{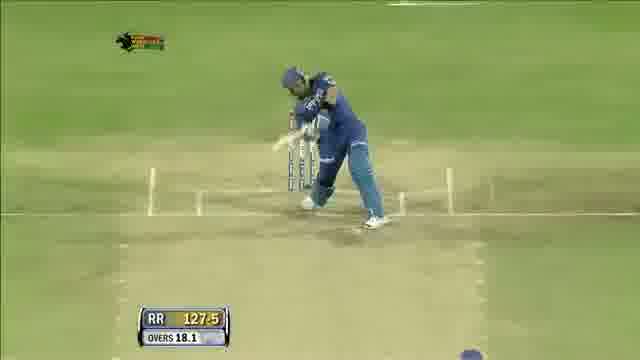} 
\includegraphics[width = 1.9cm]{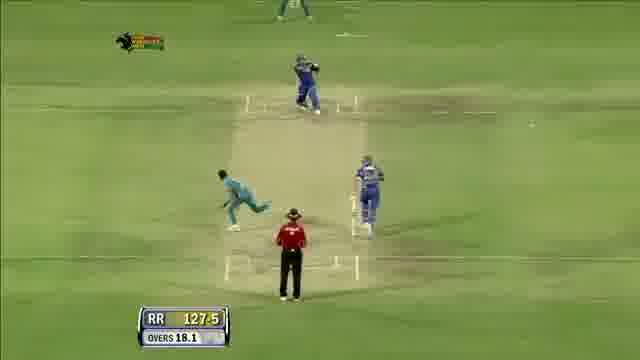} \\
\emph{that is again slammed towards long-on} \\

\hline
\vspace{0.1cm}
\includegraphics[width = 1.9cm]{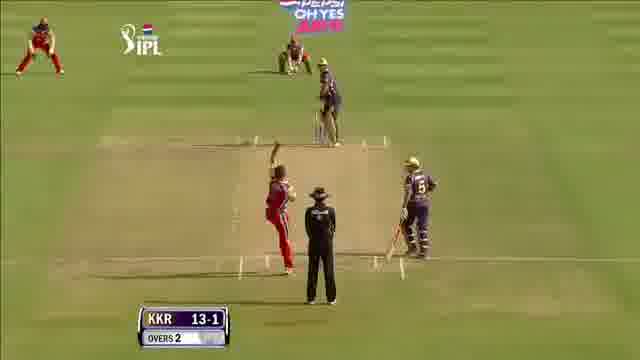} 
\includegraphics[width = 1.9cm]{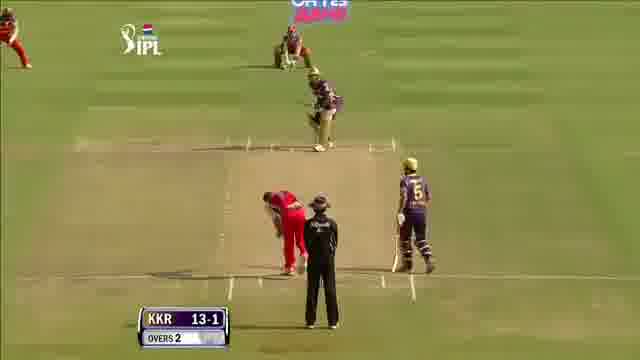} 
\includegraphics[width = 1.9cm]{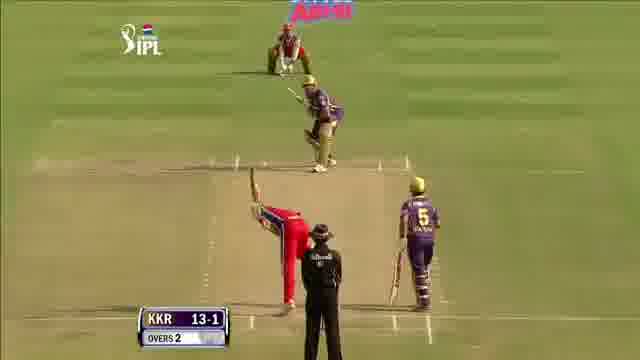} 
\includegraphics[width = 1.9cm]{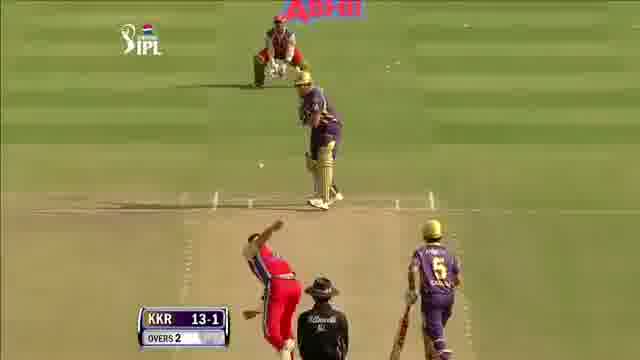} \\
\emph{shortish and swinging away lots of bounce} \\
\hline

\end{tabular}
\caption{Examples of shots correctly labeled by the batsman or bowler actions. The textual annotation is semantically rich, since it is obtained from human generated content. These annotated videos could now be used as training data for action recognition modules.}
\label{fig:qual}
\vspace{-0.6cm}
\end{figure}

\subsection{Scene Segmentation}

For the text-driven scene segmentation module, we use these mid-level features: \emph{\{Pitch, Ground, Sky, Player-Closeup, Scorecard\}}. These features are modelled using binned color histograms. Each frame is then represented by the fraction of pixels that contain each of these concepts, the scene is thus a spatio-temporal model that accumulates these scores. 

The limitation with the DP formulation is the amount of memory available to store the DP score and indicator matrices. With our machines we are limited to performing the DP over 100K frames, which amounts to 60 scenes, or 10-\emph{overs} of the match.

The accuracy of the scene segmentation is measured as the number of video-shots that are present in the correct scene segment.
We obtain a segmentation accuracy of {\bf 83.45\%}.
Example segmentation results for two scenes are presented  in Figure~\ref{fig:sceneseg},
one can notice that the inferred scene boundaries are very close to the groundtruth.
We observe that the errors in segmentation typically occur due to events that are not modelled, such as a player injury or an extended team huddle.

\subsection{Shot Recognition}

\begin{table}
\begin{tabular}{|c|c|c|c|c|}
\hline
\bf Kernel & \bf Linear & \bf Polynomial & \bf RBF & \bf Sigmoid \\
\hline
Vocab: 300 & 78.02 & 80.15 & 81 & 77.88\\
Vocab: 1000 &  \bf 82.25 & 81.16 & 82.15 & 80.53 \\
\hline
\end{tabular}
\caption{Evaluation of the video-shot recognition accuracy. A visual vocabulary using 1000 clusters of SIFT-features yields a considerably good performance, with the Linear-SVM.}
\label{tab:shotrec}
\end{table}

The accuracy of the shot recognition using various feature representations and SVM Kernels is given in Table~\ref{tab:shotrec}.
We observe that the 1000 size vocabulary works better than 300. 
The Linear Kernel seems to suffice to learn the decision boundary, with a best-case accuracy of 82.25\%.
Refining the SVM predictions with the CRF based method yields an improved accuracy of 86.54. 
Specifically, the accuracy of the batsman/bowler shot categories is 89.68\%.

\subsection{Shot Annotation Results}

The goal of the shot annotation module is to identify the right shot within a scene that contains the bowler and batsman actions. 
As the scene segmentation might contain errors, we perform a search in a shot-neighbourhood centered on the inferred scene boundary.
We evaluate the accuracy of finding the right bowler and batsman shots within a neighbourhood region $R$ of the scene boundary, which is given in Table~\ref{tab:shotannot}. 
It was observed that 90\% of the bowler and batsman shots were correctly identified by searching within a window of 10 shots on either side of the inferred boundary. 

Once the shots are identified, the corresponding textual comments for bowler and batsman actions, are mapped to these video segments.
A few shots that were correctly annotated are shown in Figure~\ref{fig:qual}.

\begin{table}
\begin{center}
\begin{tabular}{|c|c|c|}
\hline
\bf R & Bowler Shot & Batsman Shot \\
\hline
2	& 22.15 &  39.4 \\
4 	& 43.37 & 47.6 \\
6	& 69.09 & 69.6 \\
8	& 79.94 & 80.8 \\
10  & 87.87 & 88.95 \\

\hline
\end{tabular}
\end{center}	
\caption{Evaluation of the neighbourhood of a scene boundary that needs to be searched to find the appropriate bowler and batsman shots in the video. It appears that almost 90\% of the correct shots are found within a window size of 10.}
\label{tab:shotannot}
\vspace{-0.5cm}
\end{table}

\section{Conclusions}

In this paper, we present a solution that enables rich semantic annotation of Cricket videos at a fine temporal scale. 
Our approach circumvents technical challenges in visual recognition by utilizing information from online text-commentaries.
We obtain a high annotation accuracy, as evaluated over a large video collection.
The annotated videos shall be made available for the community for benchmarking, such a rich dataset is not yet available publicly.
In future work, the obtained labelled datasets could be used to learn classifiers for fine-grain activity recognition and understanding.

{\small
\bibliographystyle{ieee}
\bibliography{cricket_acpr}
}

\end{document}